\documentclass[fleqn,usenatbib]{mnras}

\usepackage[T1]{fontenc}
\usepackage{adjustbox}
\usepackage{tablefootnote}

\DeclareRobustCommand{\VAN}[3]{#2}
\let\VANthebibliography\thebibliography
\def\thebibliography{\DeclareRobustCommand{\VAN}[3]{##3}\VANthebibliography}

\newcommand{\GG}[1]{}
\usepackage{graphicx}	% Including figure files
\usepackage{amsmath}	% Advanced maths commands
\usepackage{adjustbox}
\usepackage{arydshln}
\usepackage{longtable}
\usepackage{lscape}
\usepackage{newtxtext,newtxmath}

\title[AUDFn: UVLF direct determination and evolution from z $\sim$ 0.8--0.4]{The AstroSat UV Deep Field North: Direct determination of the UV Luminosity Function and its evolution from z $\sim$ 0.8 -- 0.4}

\author[Souradeep Bhattacharya et al.]{
Souradeep Bhattacharya,$^{1}$\thanks{E-mail: souradeep@iucaa.in}
Kanak Saha,$^{1}$
and Chayan Mondal$^{1}$
\\
$^{1}$Inter University Centre for Astronomy and Astrophysics, Ganeshkhind, Post Bag 4, Pune 411007, India
}

\date{Accepted XXX. Received YYY; in original form ZZZ}

\pubyear{2023}

\begin{document}
\label{firstpage}
\pagerange{\pageref{firstpage}--\pageref{lastpage}}
\maketitle

\begin{abstract}
We characterise the evolution of the rest-frame 1500~\AA~UV luminosity Function (UVLF) from AstroSat/UVIT F154W and N242W imaging in the Great Observatories Origins Survey North (GOODS-N) field. With deep FUV observations, we construct the UVLF for galaxies at z$<0.13$ and subsequently characterise it with a Schechter function fit. The fitted parameters are consistent with previous determinations. With deep NUV observations, we are able to construct the UVLF in seven redshift bins in the range z$\sim$0.4--0.8, with galaxies identified till $\sim$2~mag fainter than previous surveys, owing to the high angular-resolution of UVIT. The fitted Schechter function parameters are obtained for these UVLFs. At z$\sim$0.7--0.8, we also utilise \textit{Hubble Space Telescope} (HST) F275W observations in the GOODS-N field to construct the UVLF in 2 redshift bins, whose fitted Schechter function parameters are then found to be consistent with that determined from UVIT at z$\sim$0.75. We thus probe the variation of the fitted UVLF parameters over z$\sim$0.8--0.4, a span of $\sim$2.7~Gyr in age. We find that the slope of the Schechter function, $\alpha$, is at its steepest at z$\sim$0.65, implying highest star-formation at this instant with galaxies being relatively more passive before and after this time. We infer that this is a short-lived instance of increased cosmic star-formation in the GOODS-N field even though cosmic star-formation
may be winding-down over longer timespan at this redshift range.
\end{abstract}

\begin{keywords}
galaxies: star formation -- ultraviolet: galaxies -- galaxies: luminosity function -- galaxies: evolution 
\end{keywords}

%%%%%%%%%%%%%%%%%%%%%%%%%%%%%%%%%%%%%%%%%%%%%%%%%%

%%%%%%%%%%%%%%%%% BODY OF PAPER %%%%%%%%%%%%%%%%%%

\section{Introduction}
\label{sec:intro}
Galaxy luminosity functions are powerful tools for understanding the properties and evolution of galaxy populations (see review by \citealt{Johnston11}). These functions quantify the distribution of luminosities within a given population, providing insights into the underlying physical processes involved in galaxy formation and evolution. The rest-frame 1500~\AA~UV luminosity function (UVLF) is of particular interest as it is sensitive to the ongoing star formation activity in galaxies \citep{wyder05}. The UV emission predominantly arises from the young, massive stars in galaxies, making it an excellent tracer of recent star formation rates (e.g. \citealt{FloresVelazquez21}). By studying the UVLF, we can probe the formation and evolution of star-forming galaxies across cosmic history.

Previous studies have extensively characterized the UVLF over a range of redshifts-- z$<1$ predominantly from \textit{Galaxy evolution Explorer} (GALEX) observations \citep{Arnouts05,wyder05}; z$\sim$1--3 from \textit{Hubble Space Telescope} (HST) surveys in deep-fields\footnote{Deep fields, the first of which was surveyed by HST \citep{Williams96}, are sensitive observations over a small region of the sky that provide unique opportunities to detect and analyze faint, low-luminosity galaxies that may contribute significantly to the overall population and constrain the faint-end of the UVLF.} \citep[e.g.][]{oesch10, Alavi16} and wide-field ground-based optical surveys \citep[e.g.][]{Cucciati12,Moutard20}; z$\sim$2--9 also from HST surveys in deep-fields \citep[e.g.][]{Bouwens15,Bouwens21}; and z$>$8 from recent \textit{James Webb Space Telescope} (JWST) observations of deep-fields \citep{Bouwens23}. These investigations have unveiled the rapid rise of star formation activity from early times till z$\sim$2, a period termed as cosmic noon, and the subsequent reduction in star-formation thereafter till present times \citep[e.g.][]{oesch10, Moutard20}. 

Direct determination of the UVLF characteristics is generally carried out by constructing the UVLF in bins of different redshifts \citep[e.g.][]{oesch10}. While at high-redshifts, these UVLFs in redshift bins represent a small range of galaxy ages ($\sim$~1.1 Gyr at z=2--3), the same range in redshift at lower z represent a wider range of galaxy ages ($\sim$~2.5 Gyr at z=1--2; $\sim$~6.5 Gyr at z=0.1--1). The near-UV (NUV) and optical observations of galaxies, predominantly from HST studies \citep[e.g.][]{oesch10} have characterised the UVLF at z$>$1 with relatively higher resolution in age. To establish a comprehensive understanding of galaxy evolution, it is crucial to extend the investigation of the UVLF to lower redshifts ($z<1$) with similar resolution in age. This redshift range offers insights into the transition from the vigorous star-forming galaxies observed at higher redshifts to the quiescent, more evolved population found in the local universe \citep[e.g.][]{Leitner12}. Understanding the evolution of the UVLF across this transition is essential for constructing a coherent picture of galaxy evolution.

GALEX FUV observations directly probe the rest-frame 1500~\AA~UVLF for very nearby galaxies (z$\sim0.1$) from wide-field surveys \citep{Arnouts05,wyder05} while GALEX NUV observations have been utilised for constructing the UVLF at z $\sim$ 0.2--1 \citep{Arnouts05}. Given the low angular resolution of GALEX (FWHM$\rm_{NUV}=5.3^{\prime\prime}$), the UVLF was not well-fitted in the faint-end and redshift bins were kept relatively larger (bin size $\sim0.2$ in z) to compensate. Few other studies have also directly determined the UVLF at the relatively higher-end of this redshift range \citep[e.g.][]{oesch10,Page21}, but none have constrained the UVLF at faint magnitude limits. While the most star-forming galaxies at any given redshift occupy the brighter end of the UVLF, the relatively more numerous passive galaxies occupy its fainter end. To well characterize the star-formation at any given redshift and in-particular the UVLF slope (discussed later in Section~\ref{sec:uvlf}), observations of the galaxies at the faint-end of the UVLF are needed.

The UV Imaging Telescope (UVIT; \citealt{Kumar12}) on board \textit{AstroSat} \citep{singh14} has carried out high angular-resolution (FWHM$\rm_{F154W}=1.18^{\prime\prime}$, FWHM$\rm_{N242W}=1.11^{\prime\prime}$; \citealt{Mondal23}) UV observations of the Great Observatories Origins Deep Survey - North and South fields (GOODS-N and GOODS-S respectively; \citealt{Giavalisco04}). The observations of the GOODS-N field are described in \citet{Mondal23} while that of the GOODS-S field will be presented in Saha et al. (2023, in preparation). These observations have already resulted in direct detections of Lyman-continuum leakers at z$\sim$1.5 \citep{Saha20,Dhiwar24} as well as the discovery of extended-UV discs around blue-compact-dwarf galaxies at z$\sim$0.2 \citep{Borgohain22} in the GOODS-S field,  and characterization of the UV continuum slope ($\rm\beta$) in the GOODS-N field \citep{Mondal23b}. 

Given the sensitivity to the 1500~\AA~region of galaxies at z$\sim0.1$ and z$\sim$0.8--0.4 from UVIT FUV and NUV imaging respectively, and the availability of well-determined photometric and spectroscopic redshifts in the aforementioned deep fields, we are finely poised to utilize these observations to construct deep UVLFs with high resolution in look-back time (with narrow redshift bins). We can obtain the UVLF parameters and construct a coherent picture of galaxy evolution over these redshift ranges. Additionally, the FUV 1500~\AA~flux measurement also allows for computation of the recent (<100~Myr) star-formation rate (SFR; \citealt{Kennicutt12}) in the observed galaxies in the aforementioned redshift window. 

We describe the data utilised in this work in Section~\ref{sec:data}, the constructed UVLFs in Section~\ref{sec:uvlf} and the derived FUV SFR in Section~\ref{sec:sfr}. We discuss the UVLF evolution in Section~\ref{sec:disc} and present our conclusions in Section~\ref{sec:conc}.

Throughout this paper, we use the standard cosmology ($\rm\Omega_{m}$ = 0.3, $\rm\Omega_{\Lambda}$ = 0.7 with $\rm H_{0}$ = 70 km~s$^{-1}$~Mpc$^{-1}$). 
Magnitudes are given in the AB system \citep{Oke74}.

%________________________________________

\section{Data} 
\label{sec:data}

Deep UV imaging observations of the GOODS-N field was carried out with AstroSat/UVIT (AstroSat UV Deep Field North, hereafter AUDFn), using one far-UV (FUV; F154W, 34.0 ks) and two near-UV (NUV) filters (N242W, 19.2 ks; N245M, 15.5 ks), as detailed in \citet{Mondal23}. The AUDFn imaging data covers $\sim$616 sq. arcmin, including the $\sim$158 sq. arcmin covered by the HST CANDELS survey with multiple HST filters \citep{Grogin11} and the 56.5 sq. arcmin spanned by the Hubble Deep UV Legacy Survey in the GOODS-N field with HST F275W and F336W filters \citep{Oesch18}. In this work, we utilize only those F154W and N242W sources catalogued by \citet{Mondal23} that have a single unique HST counterpart within $1.4''$\footnote{The radius of $1.4''$ is slightly larger than the FWHM of the UVIT NUV and FUV PSFs and sources having a separation beyond this radius will not appear blended even if there is a large magnitude difference between the sources \citep{Mondal23}.}. We additionally restrict ourselves to only those sources that are brighter than the 50\% completeness limits of 26.40 and 27.05~mag for the F154W and N242W images respectively

The redshifts for our galaxies were adopted from the z$\rm_{best}$ value of the CANDELS Photometric Redshift Catalog \citep{Kodra23}. z$\rm_{best}$ is their best estimate of a galaxy's redshift determined from the combined data set of spectroscopic redshifts, 3D-HST grism redshifts, and best photometric redshift estimate (see \citealt{Kodra23} for details). As per the filter transmission functions for the F154W and N242W filters \citep{Tandon20}, they are most sensitive to the redshifted 1500~\AA~ for galaxies at z$<=0.13$ and $0.378<$~z~$<0.768$ respectively. We thus select 95 and 1258 galaxies with only a single HST counterpart in the F154W and N242W images respectively within these redshift ranges. These are then utilised to construct the rest-frame 1500~\AA~UVLFs (see Section~\ref{sec:uvlf}). These selected galaxies survey an area of $\sim$142 sq. arcmin on the sky. 

Furthermore, the HST F275W (given its filter transmission) reliably traces the redshifted 1500~\AA~ for galaxies at $0.679<$~z~$<0.79$. There are 326 galaxies in this redshift range observed in F275W filter down to the 5$\sigma$ detection limit of 27.4~mag. As this overlaps with the higher redshift range probed by N242W, we utilise the HST F275W photometric catalogue \citep{Oesch18} to construct UVLFs as a consistency check to the N242W based UVLFs (see Section~\ref{sec:uvlf}).

%_____________________________________________
\section{The UV (1500~\AA) Luminosity Functions} 
\label{sec:uvlf}

\begin{table*}
\caption{Fitted UVLF Parameters. Column 1: Central redshift of each bin; Column 2: Range of redshifts spanned by each bin; Column 3: No. of galaxies in each bin; Columns 4--6: Fitted Schechter function parameters; Column 7: Logarithm of luminosity density.}
\centering
\adjustbox{max width=\textwidth}{
\begin{tabular}{cccccc}
\hline
$\rm z_{mean}$ & $\rm z_{range}$ & No. of & $\phi_*$ & M* &
$\alpha$ \\
 &  & sources & $10^{-3}$ mag$^{-1}$ Mpc$^{-3}$ & mag & \\
\hline
\multicolumn{6}{c}{AUDFn FUV}\\
\hline
0.1 & 0.01 -- 0.13 & 95 & $ 2.16 \pm 1.3 $ & $ -18.03 $ & $ -1.32 \pm 0.15 $\\
\hline
\multicolumn{6}{c}{AUDFn NUV}\\
\hline
0.41 & 0.378 -- 0.434 & 120 & $ 5.92 \pm 3.07 $ & $ -17.6 \pm 0.4 $ & $ -1.17 \pm 0.27 $\\
0.46 & 0.434 -- 0.489 & 243 & $ 13.45 \pm 2.34 $ & $ -18.04 \pm 0.16 $ & $ -0.84 \pm 0.14 $\\
0.52 & 0.489 -- 0.545 & 226 & $ 9.71 \pm 1.6 $ & $ -18.23 \pm 0.15 $ & $ -0.97 \pm 0.1 $\\
0.57 & 0.545 -- 0.601 & 193 & $ 3.88 \pm 1.3 $ & $ -18.77 \pm 0.25 $ & $ -1.41 \pm 0.12 $\\
0.63 & 0.601 -- 0.657 & 172 & $ 1.35 \pm 1.04 $ & $ -19.56 \pm 0.59 $ & $ -1.65 \pm 0.17 $\\
0.68 & 0.657 -- 0.712 & 153 & $ 2.53 \pm 0.95 $ & $ -19.29 \pm 0.31 $ & $ -1.33 \pm 0.13 $\\
0.74 & 0.712 -- 0.768 & 151 & $ 6.11 \pm 3.2 $ & $ -18.45 \pm 0.44 $ & $ -0.97 \pm 0.43 $\\
\hline
\multicolumn{6}{c}{HDUV-N}\\
\hline
0.71 & 0.679 -- 0.734 & 143 & $ 2.75 \pm 1.3 $ & $ -18.44 \pm 0.45 $ & $ -1.07 \pm 0.2 $\\
0.76 & 0.734 -- 0.79 & 183 & $ 3.45 \pm 1.73 $ & $ -18.66 \pm 0.43 $ & $ -1.11 \pm 0.24 $\\
\hline
\end{tabular}
\label{tab:uvlf}
}
\end{table*}

%_____________________________________________

\begin{figure}
\includegraphics[width=\columnwidth]{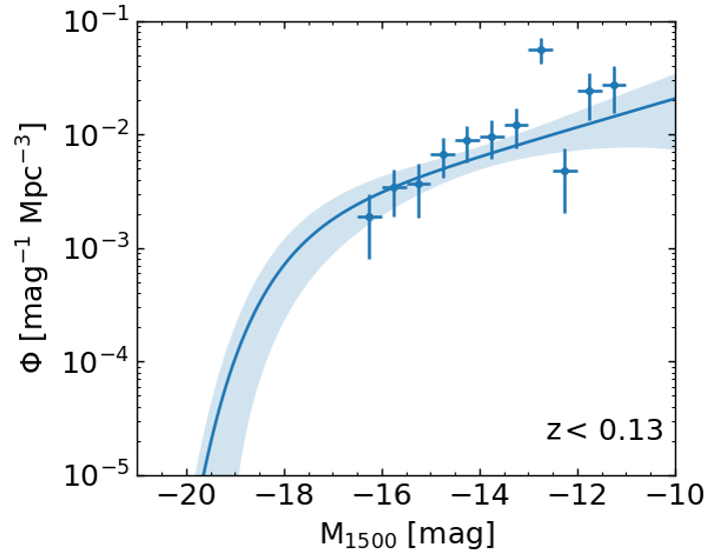}
\caption{The 1500~\AA~LF at z~$<0.13$ from the F154W imaging of the GOODS-N field. The Schechter function fit is marked with a solid blue line, with the 1~$\rm\sigma$ uncertainty shaded.}
\label{fig:fuv}
\end{figure}

\begin{figure*}
\includegraphics[width=\textwidth]{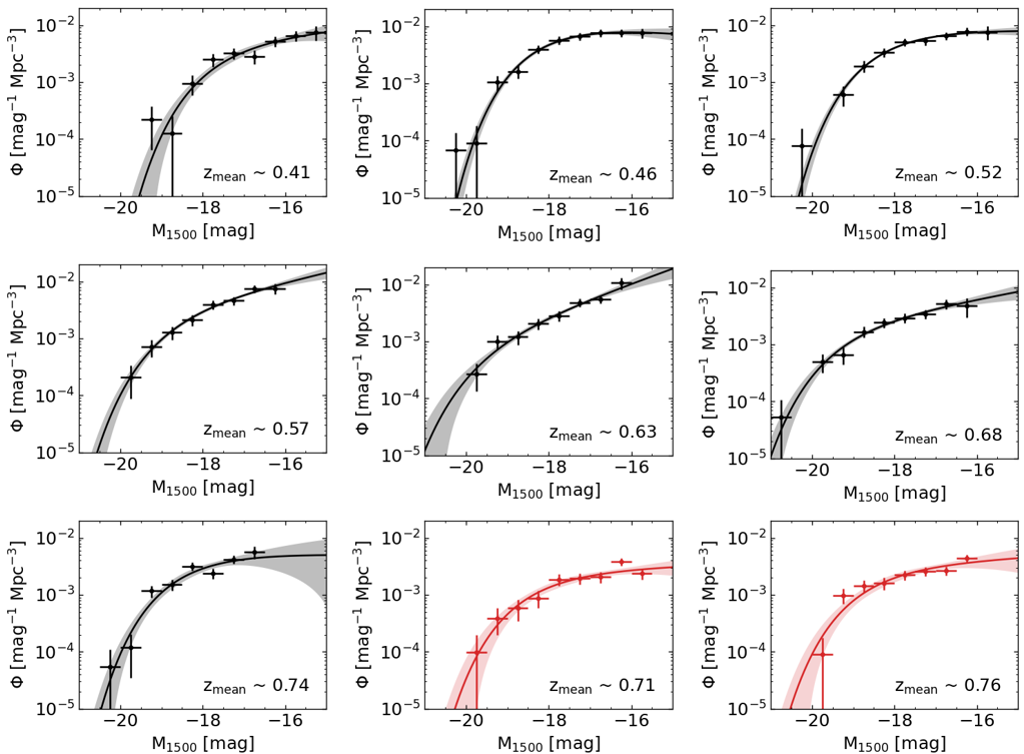}
\caption{The 1500~\AA~LF from AstroSat/N242W images of the GOODS-N field in seven different redshift bins are marked in black. The z$\rm_{mean}$ of each bin is also noted. The Schechter function fits for each bin is marked with a solid black line, with the 1~$\rm\sigma$ uncertainty shaded in grey. The center and right panels of the bottom row show the 1500~\AA~LF from HST F275W images of the GOODS-N field in two different redshift bins, marked in red. The Schechter function fit for each bin is marked with a solid red line, with the 1~$\rm\sigma$ uncertainty shaded. }
\label{fig:uvlf}
\end{figure*}

\begin{figure*}
\includegraphics[width=\textwidth]{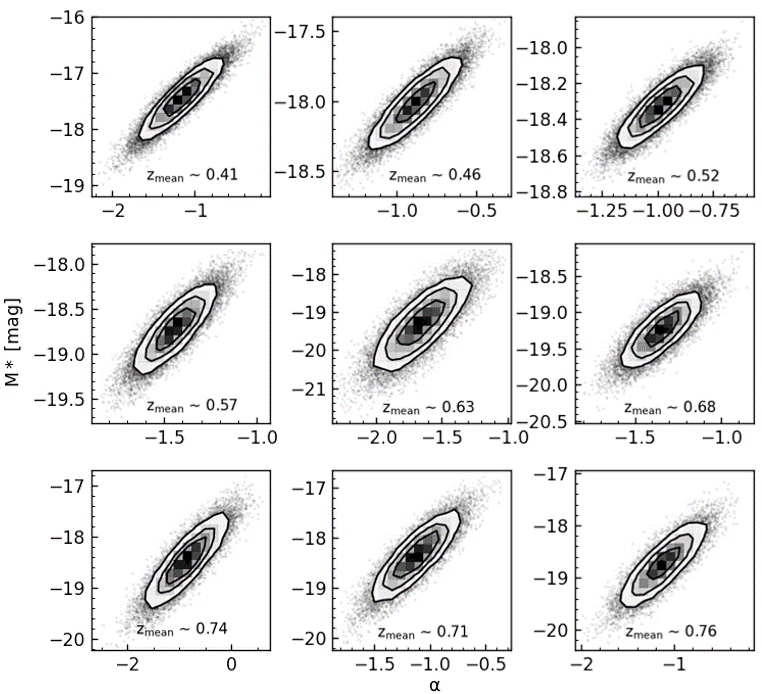}
\caption{Each panel shows the correlation between fitted 
$\alpha$ and M* of the Schechter function fit (presented as in Figure~\ref{fig:uvlf}) for each redshift bin. The z$\rm_{mean}$ of each bin is also noted. The 1, 2 \& 3 $\sigma$ confidence intervals are marked by the innermost, middle and outermost contours in each panel. The possible solutions within 3 $\sigma$ are binned while those beyond are marked as individual points. The first seven and last two panels are for the AstroSat/UVIT N242W and HST/HDUV F275W UVLFs respectively.}
\label{fig:correlation}
\end{figure*}

\subsection{The AstroSat/UVIT F154W Luminosity Function} 
\label{sec:fuv_uvlf}
The measured F154W magnitudes of the 95 galaxies are utilised to construct the UVLF at z$<=0.13$. Completeness correction is based on the recovery fraction of sources as a function of F154W magnitude (see Figure~14 in \citealt{Mondal23}) that was carried out using artificial source injection \citep{Bhattacharya19} for the F154W image. A further correction is applied to account for the fraction of total identified sources that have a unique HST counterpart. The small K-corrections are derived from appropriate best-fit templates. 

Accounting for the aforementioned corrections and the luminosity distance of each galaxy, we construct the UVLF shown in Figure~\ref{fig:fuv}. The binned representation of the UVLF follows the method of \citet{Page00} and clearly shows in Figure~\ref{fig:fuv} that galaxies are observed down to M$_{1500}$=-11~mag. The UVLF is then fitted with the classic Schechter function \citep{Schechter76} using a maximum likelihood estimator fit, similar to that carried out by \citet{Page21}. The Schechter function can be parameterised as a function of
magnitude M with parameters $\alpha$, $\phi_{*}$ and M$^{*}$ as follows:
\begin{equation}
    \phi(M) dM = \frac{ln(10)}{2.5}~\phi_{*}~(10^{0.4\Delta M})^{\alpha + 1}e^{-10^{0.4\Delta M}}dM
\end{equation}
where $\Delta M = M^* - M$. Here, $\phi_{*}$ is  a normalization factor which defines the overall density of galaxies (number per cubic Mpc).  $M^*$ is the characteristic galaxy magnitude. An $M^*$ galaxy is a bright galaxy, roughly
comparable in luminosity to the Milky Way. $\alpha$ defines the faint-end slope of the UVLF, which is typically negative
implying large numbers of relatively passive galaxies with low luminosities.

As there is a clear paucity of bright sources at this redshift range in the AUDFn field, we fix $M^*=-18.03$, the same value as that obtained by \citet{Arnouts05} using GALEX FUV data. The fitted $\phi_{*}$ and M$^{*}$ values thus obtained are noted in Table~\ref{tab:uvlf}. 

%_____________________________________________

\subsection{The AstroSat/UVIT N242W Luminosity Functions} 
\label{sec:nuv_uvlf}
The measured N242W magnitudes of the 1258 galaxies are utilised to construct the UVLFs at $0.378<$~z~$<0.768$. These galaxies are divided into 7 equal bins in redshift ($\rm\Delta z = 0.055$) such that at least 100 galaxies are present in each bin, with at least 1 galaxy populating the bright end of the UVLF for each bin. The mean redshift of each bin, z$\rm_{mean}$, the range of redshifts as well as the number of galaxies in each bin are noted in Table~\ref{tab:uvlf}. 

For the galaxies in each redshift bin, corrections are applied similar to that described in Section~\ref{sec:fuv_uvlf}. The binned representation of the UVLF for each redshift bin clearly shows in Figure~\ref{fig:uvlf} that galaxies are observed down to M$_{1500}$=-15~mag at z$\rm_{mean}$=0.41, but this limit decreases with increased z$\rm_{mean}$ values and galaxies in the UVLF are observed only down to M$_{1500}$=-16.5~mag at z$\rm_{mean}$=0.74. The Schechter function is fitted to the UVLF (as described in Section~\ref{sec:fuv_uvlf}) in each redshift bin keeping $\alpha$, $\phi_{*}$ and M$^{*}$ as free parameters. The best-fit Schechter function parameters are noted in Table~\ref{tab:uvlf}. The 1~$\sigma$ shaded uncertainty to the best-fit Schechter function is plotted in Figure~\ref{fig:uvlf}, showing that the UVLFs are well-constrained over $\sim4$~mag below the brightest M$_{1500}$ value in each redshift bin. 

We note also that the best-fit Schechter function parameters are correlated, as has been found also in previous studies of the UVLF (E.g. \citealt{Iwata07,oesch10}). In particular, we show the covariance between $\alpha$ and M* in Figure~\ref{fig:correlation} for each redshift bin. Through Monte-Carlo resampling of the UVLFs, we have obtained the 1, 2 \& 3 $\sigma$ confidence intervals. The correlation between $\alpha$ and M* is similar in the different redshift bins, as evident from the similar area spanned by the 1, 2 \& 3 $\sigma$ contours for each redshift bin. The 1 $\sigma$ errors for the Schechter function parameters are noted in Table~\ref{tab:uvlf}.

\subsection{AGN contribution to the AstroSat/UVIT N242W Luminosity Functions} 
\label{sec:nuv_uvlf_agn}

At high redshift (z$\sim$5--6), a substantial contribution of active galactic nuclei (AGN) to the observed UVLF is expected from theoretical models \citep{Piana22}, which may be more prominent at z$\sim$3.5--4 \citep{Trebitsch21}. A comparison with known AGN identified in the 2 Ms Chandra Deep Field-North Survey \citep{Xue16} results in only 26 AGN in common with our 1258 sources in AUDFn within the redshift range of interest ($0.378<$~z~$<0.768$). These AGN cover M$_{1500}=-19.08$~--~$-15.66$~mag, with mean M$_{1500}=-17.7$~mag and $\sigma$(M$_{1500})=0.99$~mag. There are 1--6 AGN in each redshift bin, with the highest AGN occuring in the z$\rm_{mean}$=~0.57~\&~0.63 bins and the lowest in the z$\rm_{mean}$=0.68 bin. Only three AGN have M$_{1500}>-19$~mag, occuring in the z$\rm_{mean}$=~0.57,~0.63~\&~0.74 bins. Removing all the known AGN from our sample only changes the fitted UVLF parameters $\alpha$, $\phi_{*}$ and M$^{*}$ by 0.5--3\%, 0.8--8\% and 0--0.3\% respectively. The z$\rm_{mean}$=~0.57~\&~0.63 bins with the highest number of AGN show the maximum change of $\phi_{*}$ by 5.13~\&~8\% respectively. For M$^{*}$, the maximum change of 0.3\% occurs in the z$\rm_{mean}$=~0.57 bin, which is the lowest redshift bin with an AGN having M$_{1500}>-19$~mag. There is no clear correspondence of the change in $\alpha$ with the presence of the maximum number or the brightest AGN. In any case, the removal of AGN only changes the best-fit UVLF parameters well within their
estimated uncertainties (see Table~\ref{tab:uvlf}). Unlike the expectations for higher redshift galaxies, UVLF parameters thus appear to have negligible effect due to AGN at $0.378<$~z~$<0.768$.

%_____________________________________________
\subsection{The HST/HDUV F275W Luminosity Functions} 
\label{sec:hst_uvlf}
The measured F275W magnitudes of the 326 galaxies catalogued by \citet{Oesch18} are utilised to construct the UVLFs at $0.679<$~z~$<0.79$. The galaxies are divided into two redshift bins ($\rm\Delta z = 0.055$) following the criteria described in Section~\ref{sec:nuv_uvlf}. The binned representation of the UVLF is then constructed, as shown in red in Figure~\ref{fig:uvlf}, without any completeness correction\footnote{\citet{Oesch18} had constructed the catalogue of HDUV galaxies based on detection in NIR HST filters. The completeness in their F275W catalogue is thus based on the detection completeness in these NIR filters. As these galaxies show a spread in F275W - F160W colour over $\sim6$~mag, the detection completeness in  NIR filters are not an accurate representation of the completeness in F275W. However, given the purpose of comparing the UVLF fit parameters to that obtained from the AstroSat/N242W observations in the same redshift range, we construct the F275W UVLFs without any completeness correction, understanding that the effect of the completeness correction will be mild as we restrict ourselves to the 5$\sigma$ detection limit, that is generally expected to be brighter than the 50\% completeness limit.}. Galaxies are observed down to M$_{1500}$=-16.5~mag and -16~mag at z$\rm_{mean}$=0.71 and 0.76 respectively, deeper than in the N242W data at the same redshift. The Schechter function is then fitted to the UVLF (as described in Section~\ref{sec:fuv_uvlf}) in both redshift bins keeping $\alpha$, $\phi_{*}$ and M$^{*}$ as free parameters. The best-fit Schechter function parameters are noted in Table~\ref{tab:uvlf} and show that the fitted M$^{*}$ values are similar (with similar uncertainty) for the HDUV UVLFs and the N242W UVLF at z$\rm_{mean}$=0.71. The fitted $\alpha$ and $\phi_{*}$ values also agree within errors but the uncertainty is lower (by almost 50\%) for the deeper HDUV UVLFs. Thus the directly determined HDUV UVLFs and the N242W UVLF for the GOODS-N field are consistent for the overlapping redshift range covered by the two filters. We note that the covariance between $\alpha$ and M* for the two HST/HDUV F275W UVLFs is also shown in Figure~\ref{fig:correlation}.

%_____________________________________________

\begin{figure}
\includegraphics[width=\columnwidth]{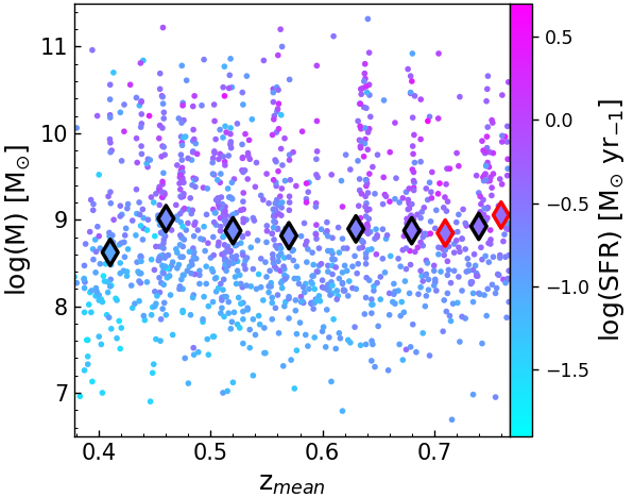}
\caption{The logarithm of the stellar mass of each galaxy \citep{Barro19} with N242W observations at $0.378<$~z~$<0.768$ is plotted against their redshift, coloured by the logarithm of their computed FUV SFR. Marked as diamonds are the mean log(M), coloured by mean log(SFR), at the z$\rm_{mean}$ values corresponding to the 7 bins where UVLFs are computed for AUDFn NUV observations (black outline) and the 2 bins where UVLFs are computed for HDUV-N observatons (red outline).}
\label{fig:z_vs_mass}
\end{figure}

\section{Galaxy stellar mass and FUV star formation from z$\sim$0.8--0.4} 
\label{sec:sfr}

While a number of different observational methods exist to determine the SFR of a galaxy \citep[e.g.][]{Popesso23}, the measured FUV 1500~\AA~flux may be utilised to compute the FUV 1500~\AA~luminosity and thereafter derive the dust-unobscured FUV SFR\footnote{We are not correcting for internal extinction in this work. As FUV flux is generally heavily obscured by dust, so our estimated SFR is a lower limit of the total SFR. We obtain this dust-unobscured FUV SFR following Equation 3 in \citet{Murphy11}:
\begin{equation}
    $$ \rm SFR_{FUV}~[M_{\odot}~ yr^{-1}] = 4.42 \times 10^{-44}~L_{FUV}~[erg~s^{-1}] $$ 
\end{equation}} (pertaining to the star-formation over the past $\sim$10--100~Myr; \citealt{Kennicutt12}). We thus calculate the SFR for the 1258 galaxies with N242W observations at $0.378<$~z~$<0.768$ and also for the 326 galaxies (as in Section~\ref{sec:nuv_uvlf}) at $0.679<$~z~$<0.79$ with F275W observations (as in Section~\ref{sec:hst_uvlf}). The stellar masses (M) of these galaxies had already been computed by \citet{Barro19} from stellar population model fits to their spectral energy distributions from multi-wavelength photometric observations.

Figure~\ref{fig:z_vs_mass} shows the variation of stellar masses of each galaxy with N242W observations as a function of redshift, coloured by the logarithm of their computed FUV SFR. As we are comparing only amongst star-forming galaxies, i.e., only those having any rest-frame FUV emission, we find that the more massive galaxies naturally have higher SFR than lower mass ones (following the main sequence of star-forming galaxies; \citealt{Noeske07}). The mean stellar mass of galaxies in each redshift bin, where the UVLF was computed, is also plotted. We find a mean stellar mass of galaxies in each bin $\sim10^{9}$~M$_{\odot}$, except the lowest redshift bin which seems to have a lower mean stellar mass. 

\begin{figure}
\includegraphics[width=\columnwidth]{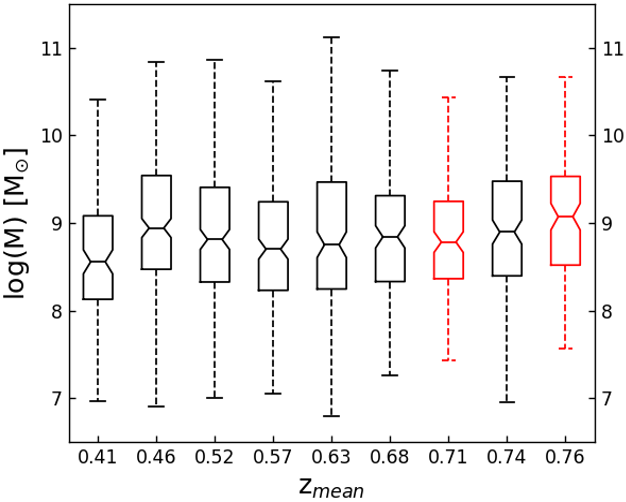}
\caption{Box-Whisker plot showing the characteristics of the log(M) distributions of galaxies in the different redshift bins studied in this work (AUDFn NUV : black; HDUV-N : red). For each box, the line in the center is the median value of log(M), while the top and bottom edges of the box represent the upper and lower quartiles respectively; the values at which the top and bottom horizontal lines stop are the minimum and maximum values of log(M).}
\label{fig:mass_box}
\end{figure}

To better visualize the stellar mass distribution of galaxies in each redshift bin, we show box-whisker plots of log(M) as a function of z$\rm_{mean}$ in Figure~\ref{fig:mass_box}. The median values of log(M) does not systematically vary with redshift and neither does the inter-quartile range (the box; which accounts for 50\% of the galaxy stellar masses in each redhift bin), which seems to overlap for the different redshift bins though some variation is notable. The minimum to maximum stellar mass range (the whiskers) also sees some variation with redshift but they also overlap over the whole range and show no trend with redshift. 

To further determine any change in the stellar mass distribution of the galaxy sample as a function of redshift, we carry out a statistical comparison of the log(M) distributions in pairs of redshift bins with an Anderson-Darling test (AD-test\footnote{The AD-test usage has been discussed in detail in \citet[][see Section~3.4]{Bh21} along with applications in \citet{Bhattacharya22,Bhattacharya22a,Bhattacharya23}. It is adopted here to allow the comparison of numerically different data-sets.}; \citealt{ADtest}). Comparing the log(M) distribution at z$\rm_{mean}=0.41$ with that at z$\rm_{mean}=0.46$, we obtain a test significance (equivalent to a p-value) of 0.001. Since the significance is below 0.05, the two log(M) distributions identified as being drawn from different parent distributions. Comparing the log(M) distribution at z$\rm_{mean}=0.46$ with that at z$\rm_{mean}=0.52$, that at z$\rm_{mean}=0.52$ with that at z$\rm_{mean}=0.57$, and so on, we find no other pair with significance below 0.05, i.e, all other pairs of log(M) distributions with increasing redshift may or may not be drawn from different parent distributions.

We, thus, statistically find that the galaxies in the redshift bin with z$\rm_{mean}=0.41$ have lower log(M) than the galaxies in the higher redshift bins. Galaxies in other redshift bins do not have any statistically different stellar masses.

\begin{figure*}
\includegraphics[width=\textwidth]{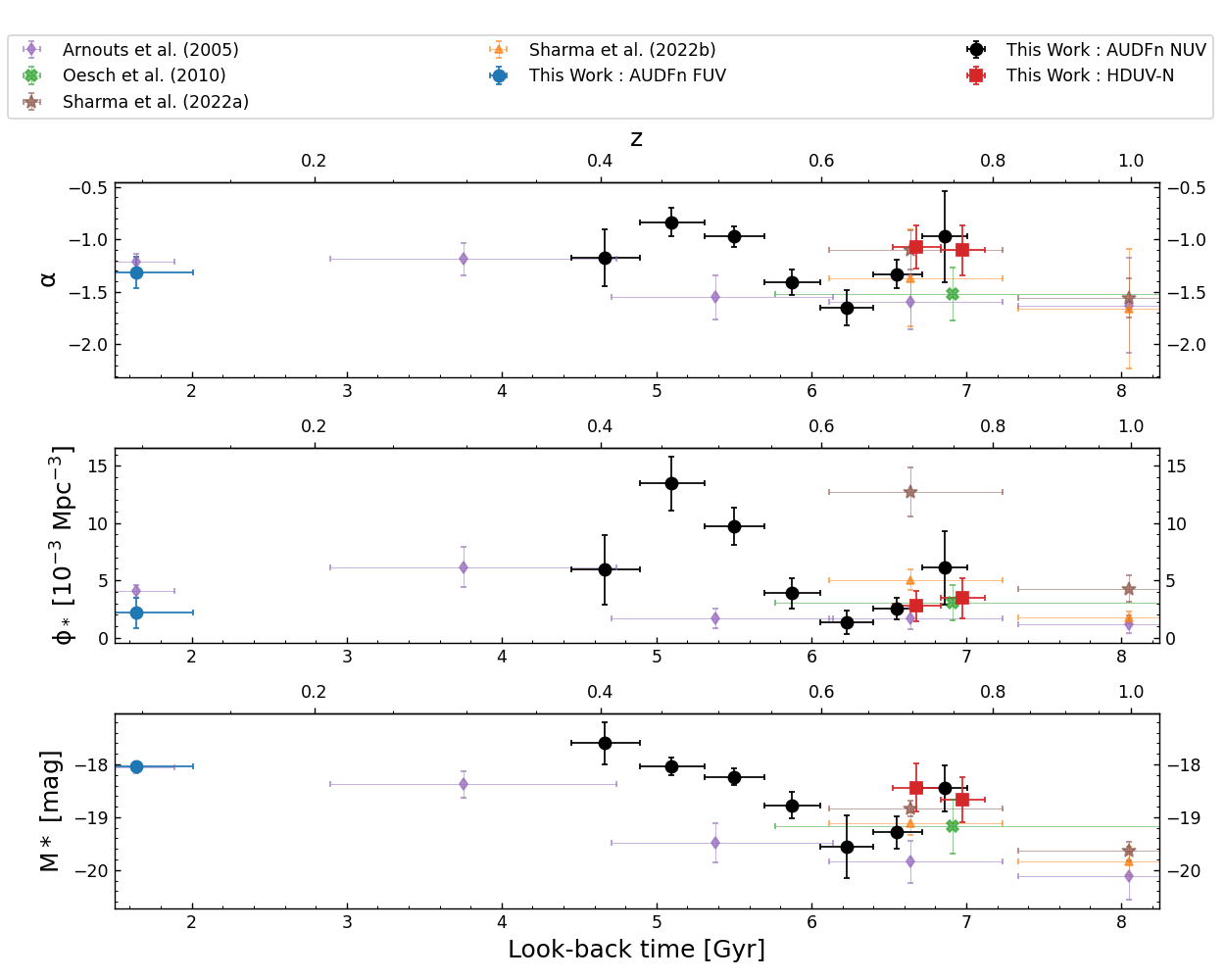}
\caption{Evolution of the fitted Schechter function parameters with look-back time (bottom axes) and redshift (top axes). The parameters from this work and from literature estimates are marked separately.} 
\label{fig:param}
\end{figure*}

%_____________________________________________
\section{Discussion} 
\label{sec:disc}
%_____________________________________________
\subsection{Consistency with past determinations of the UVLF} 
\label{sec:lit}

Figure~\ref{fig:param} shows the variation of the three Schechter function parameters, $\alpha$, $\phi_{*}$ and M$^{*}$, with redshift in this work, as well as other direct determinations of these parameters up to z$\sim1$ in the literature. At the lowest redshift (z$\sim0.1$) for the UVLF from AstroSat/F154W,  M$^{*}$ was kept constant to the value determined by \citet{Arnouts05} from GALEX FUV imaging. The $\alpha$ and $\phi_{*}$ values for the best fit F154W UVLF are also consistent with the values determined by \citet{Arnouts05}. 

From GALEX NUV imaging, \citet{Arnouts05} also determined the UVLF fit parameters for z=0.2--0.4, 0.4--0.6, 0.6--0.8 and 0.8--1.2, which are plotted as magenta diamonds in Figure~\ref{fig:param}. Their determinations were restricted to fitting the UVLF down to M$_{1500}\sim$-15.5, -17, -18 and -19~mag for the aforementioned respective redshift bins.  \citet{oesch10} determined the UVLF fit parameters in various redshift bins from HST F225W, F275W and F336W imaging in the GOODS-S field covering z$\sim$0.5--2.5. Their fitted parameters for the UVLF (restricted down to M$_{1500}\sim$-17) at z=0.5--1 are plotted as green crosses in  Figure~\ref{fig:param}. XMM-Newton Optical Monitor (XMM-OM) telescope images have also been utilised to obtain the UVLF fit parameters for z= 0.6--0.8 and 0.8--1.2 \citep{Page21,Sharma22a,Sharma22b}. We show the determined parameters in Figure~\ref{fig:param} for the XMM-OM UVLF determinations in the GOODS-S \citep[][in brown]{Sharma22a} and COSMOS \citep[][in orange]{Sharma22b} fields. In the z= 0.6--0.8 bin, their determinations were restricted to fitting the UVLF down to M$_{1500}\sim$-17.5 and -19~mag for the two fields respectively. 

Indirect determinations of the UVLF at this redshift range has been carried out by \citet{Cucciati12} using spectral energy distribution (SED) fits to higher wavelength images of galaxies using stellar population models and by \citet{Moutard20} using UVLFs determined from U-band images of galaxies normalized to shallower GALEX FUV image based UVLFs. The UVLF fit parameters determined in these studies also span the same parameter space as those marked in Figure~\ref{fig:param} but show nearly constant values of the parameters over the determined entire redshift range, a likely consequence of the assumptions of these indirect determination techniques (see \citealt{Moutard20} for details).

As clearly seen from Figure~\ref{fig:param}, the determination of the UVLF parameters has been carried out in this work (from AstroSat/N242W and HDUV surveys) with a significantly higher resolution in redshift than previous studies with a similar level of uncertainty, allowing for UVLF fits in smaller redshift bins. Additionally, we fit the UVLF down to fainter M$_{1500}$ values. For z=0.2--0.4, the fitted UVLF parameters from \citet{Arnouts05} are consistent within error for that determined at z$\rm_{mean}$=0.41 in this work. For z=0.4--0.6, where they fit UVLFs to shallower data, we determine higher values for $\alpha$, $\phi_{*}$ and M$^{*}$. At z=0.6--0.8, determinations from fitting the UVLF to relatively shallower data from different works \citep{Arnouts05,oesch10,Sharma22a,Sharma22b}, still results in fitted UVLF parameters that agree within error, though \citet{Sharma22a} seem to find a relatively high $\phi_{*}$. The fitted UVLF parameters from this work are also consistent with these literature determinations within error.

%_____________________________________________
\subsection{Evolution of the UVLF from z$\sim$0.8--0.4} 
\label{sec:evo}

\begin{figure}
\includegraphics[width=\columnwidth]{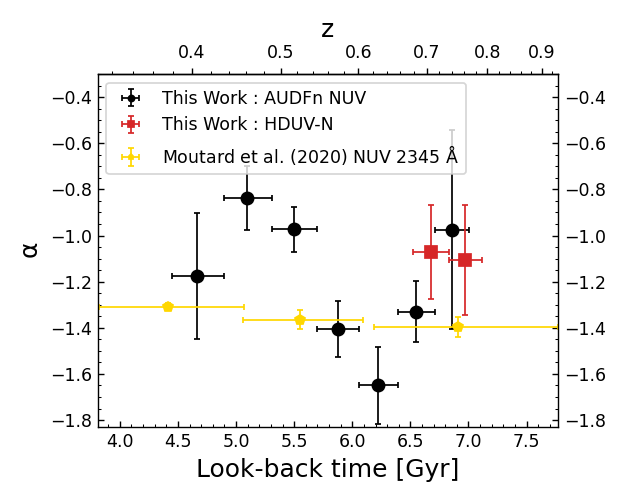}
\caption{Evolution of the fitted UVLF slope, $\alpha$ with with look-back time (bottom axis) and redshift (top axis). The rest frame 1500~\AA~UVLF slope determined in this work is compared with the rest frame NUV 2345~\AA~UVLF slope from \citet{Moutard20}. }
\label{fig:alpha}
\end{figure}

Figure~\ref{fig:param} shows the evolution of the fitted UVLF parameters from z$\sim$0.8--0.4. At z$<0.545$,  $\alpha$ is nearly constant but then declines to its lowest value at z$\rm_{mean}$=0.63 and then increases with increasing redshift. We can check if the $\alpha$ values for different redshift bins determined in this work (see Table~\ref{tab:uvlf}) may be explained by the statistical error on a single $\alpha$ value. We construct a uniform distribution with the mean and standard deviation of the determined $\alpha$ values. With a Kolmogorov–Smirnov test \citep{kstest}, we compare this uniform distribution with the $\alpha$ values. With a p-value of 0.038 (as it is $<0.5$), we reject the null hypothesis that these $\alpha$ values may be derived from single $\alpha$ value. Variations with redshift, similar to that of $\alpha$, are also seen for $\phi_{*}$ and M$^{*}$, though $\phi_{*}$ at z$\rm_{mean}$=0.41 is relatively low. As found in Section~\ref{sec:sfr}, the log(M) of galaxies at z$\rm_{mean}$=0.41 is lower than other redshift windows. This probably affects the determined Schecter function parameters at this redshift.

The $\phi_{*}$ and M$^{*}$ values determined in this work may also be affected by cosmic variance of $\sim$22\%, using the web tool by \citet{Trenti08}. However, $\alpha$ is expected to remain unaffected. As seen in Section~\ref{sec:sfr}, the stellar mass distribution of galaxies is similar in each redshift bin probed in this work. Thus the identification of a steeper $\alpha$ value of the rest-frame 1500~\AA~LF at z$\sim$0.65 is not likely an observational sample selection effect. A relatively steeper $\alpha$ value implies larger numbers of faint galaxies have high rest-frame UV emission, implying a larger number of faint star-forming galaxies. Thus, the trend in $\alpha$ seen in Figure~\ref{fig:param} may stem from a physical standpoint of star-formation increasing from z$\rm_{mean}$=0.76 to z$\rm_{mean}$=0.63 and then decreasing again till z$\rm_{mean}$=0.52 and remaining nearly constant thereafter. %Indeed in Section~\ref{sec:sfr}, we find that log(SFR) sees a noticeable transition from z$\rm_{mean}$=0.68 to z$\rm_{mean}$=0.63.

Cosmic star-formation is understood, to have peaked at z$\sim$2 (see \citealt{Forster20} and references therein) where-after star-formation has decreased till its present quiescent phase \citep[e.g.][]{Saito20}. It remains yet unclear if the cosmic star-formation winded down monotonically or there were short periods of relatively higher and lower star formation as cosmic star formation experienced its overall decline. Our identification of steeper $\alpha$ values of the rest-frame 1500~\AA~LF at z$\sim$0.65 is indicative of the latter scenario where an increase in star-formation took place over a shorter period around this redshift.

The 1500~\AA~LF is most sensitive to recent star-formation in these galaxies at short timescales between $\sim$10--100~Myr depending on the burstiness of star-formation \citep{FloresVelazquez21}. As we have smaller redshift bins (corresponding to $\sim$300~Myr), we are sensitive to small changes in star-formation with redshift. \citet{Moutard20} directly determine the rest-frame NUV (2345~\AA) UVLF from CFHT u-band and HSC g-band observations over z=0.3--0.9. As this rest-frame NUV probes the light from a relatively older stellar population, it is expected that short recent bursts would not affect their determined UVLF parameters as much as that of the rest-frame 1500~\AA~LF. 

Figure~\ref{fig:alpha} shows the fitted $\alpha$ from the 1500~\AA~LF in this work juxtaposed with the fitted $\alpha$ from the rest-frame NUV (2345~\AA) UVLF by \citet{Moutard20}. Despite the wider redshift bins, the slightly declining NUV $\alpha$ from \citet{Moutard20} shows that star-formation has been declining over longer timescales in the probed redshift range. Over shorter timescales of star-formation, as probed by the rest-frame 1500~\AA~LF based $\alpha$ values determined in this work, galaxies have been more passive than average at z=0.7--0.8, showing locally peaked star-formation at z$\sim$0.65 and then becoming more passive again at lower redshifts. The shallower $\alpha$ value for the rest-frame NUV compared to the FUV at all redshifts, other than z$\sim$0.65, also signifies that a number of fainter galaxies exist at these redshifts which have less star-formation over the past $\sim$10--100~Myr, in-line with the expectations of an overall declining cosmic star-formation history since cosmic noon. 

Probing the entire redshift range till cosmic noon with similarly high resolution in redshift, as achieved in this work, may reveal numerous short-lived instances of increasing and decreasing cosmic star-formation till its maximum at cosmic noon. On the other hand, the fluctuations in star-formation observed as function of redshift in this work, may be a local phenomenon in the AUDFn field caused by a combination of statistical fluctuations and cosmic variance. To better probe the former scenario of short-lived global fluctuations of star formation superimposed on its slow decline since cosmic noon, the UVLF needs to be determined with similarly fine high redshift bins, as achieved in this work, in other observed patches of the universe. 

%We can speculate that at z$\sim$0.65, increased star formation over the short time-scale has been driven by galaxy mergers, wherein relatively passive galaxies have experienced bursts of star-formation following in-falls of gas-rich satellites. After the newly acquired gas has been exhausted, galaxies then returned to their pre-merger star-formation rates. 

%_____________________________________________
\section{Conclusion} 
\label{sec:conc}

In this work, we present the rest-frame 1500~\AA~LF in the GOODS-N field from AstroSat/F154W (z$<$0.13), AstroSat/N242W (z$\sim$0.4--0.8) and HST/F275W (z$\sim$0.7--0.8) observations. The UVLFs thus determined are fitted with the Schechter function and the fitted parameters are examined. We find that the AstroSat/F154W based UVLF is consistent with that from GALEX at z$\sim$0.1 from their fitted Schechter parameters (Figure~\ref{fig:param}). The AstroSat/N242W based UVLFs show a significant improvement in the redshift resolution compared to literature measurements, probing z$\sim$0.4--0.8 in 7 redshift bins with similar levels of uncertainty as in previous works with wider redshift bins. The UVLFs are also fitted down to significantly fainter magnitudes. At z$\sim$0.7--0.8, the AstroSat/N242W based UVLFs are consistent with those obtained from HST/F275W observations. 

The higher redshift resolution has allowed us to probe the variation of the fitted UVLF parameters over z$\sim$0.4--0.8, a span of $\sim$2.7~Gyr in age. We find that $\alpha$ is at its steepest at z$\sim$0.65, implying highest star-formation at this time with galaxies being relatively more passive before and after this time. We infer that this is may be a short-lived instance of increased star-formation in the AUDFn field and numerous such instances may be found, by future studies with similarly fine redshift bins, as star-formation goes through peaks and troughs before reaching its maximum at cosmic noon. Such studies in other observed fields are required to determine the global nature of such instances of increased star-formation in the universe.

%_____________________________________________

\section*{Acknowledgements}

We thank the anonymous referee for their comments. SB is funded by the INSPIRE Faculty award (DST/INSPIRE/04/2020/002224), Department of Science and Technology (DST), Government of India. This work is primarily based on observations taken by AstroSat/UVIT. The UVIT project is a result of collaboration between IIA, Bengaluru, IUCAA, Pune, TIFR, Mumbai, several centres of ISRO, and CSA.  Several groups from ISAC (ISRO), Bengaluru, and IISU (ISRO), and Trivandrum have contributed to the design, fabrication, and testing of the payload. The Mission Group (ISAC) and ISTRAC (ISAC) continue to provide support by making observations with, and reception and initial processing of the data. This research made use of Astropy-- a community-developed core Python package for Astronomy \citep{Rob13}, SciPy \citep{scipy}, NumPy \citep{numpy} and Matplotlib \citep{matplotlib}. This research also made use of NASA’s Astrophysics Data System (ADS\footnote{\url{https://ui.adsabs.harvard.edu}})

%%%%%%%%%%%%%%%%%%%%%%%%%%%%%%%%%%%%%%%%%%%%%%%%%%
\section*{Data Availability}

The AUDFn source catalogue used in this work has already been published in \citet{Mondal23}. The HDUV survey catalogue has been published by \citet{Oesch18}. The CANDELS photometric redshift catalogue has been published by \citet{Kodra23}.



\bibliographystyle{mnras}
\bibliography{uvit} 

%_______________________________________________

% Don't change these lines
\bsp	% typesetting comment
\label{lastpage}
\end{document}